# Data Analysis in Social Networks for Agribusiness - A Systematic Mapping Study

Nedson D. Soares[1][0000-0003-0382-0306], Regina Braga[1][0000-0002-4888-0778], José Maria N. David[1][0000-0002-3378-015X], Kennya B. Siqueira[2][0000-0001-6727-7774], Victor Stroele[1][0000-0001-6296-8605]

nedson.soares@estudante.ufjf.br, regina.braga@ufjf.edu.br, jose.david@ufjf.edu.br, kennya.siqueira@embrapa.br, victor.stroel@ice.ufjf.br

[1] Universidade Federal de Juiz de Fora, Juiz de Fora - MG, Brasil

[2] Embrapa Gado de Leite, Juiz de Fora - MG, Brasil

**Abstract.** The ability of companies to react to changes imposed by the market is related to information acquisition and knowledge generation. Big data technologies, crowdsourcing, and Online Social Network (OSN) are used for knowledge generation. These technologies assumed a significant position in agribusiness. This work investigates how social network analysis can promote agribusiness to provide a basis for future applications and evaluations. We adopted a hybrid systematic mapping to conduct the investigation. Two hundred twenty-three works that propose solutions for agribusiness were found and categorized. Results showed the most used techniques, OSNs, and revealed an increase in the number of studies in this area. The information obtained indicates how social media monitoring can complement traditional methods for decision-making on the management and regulation of agricultural systems. However, agribusiness still lacks more studies using data analysis tools on social networks. Based on our results, we discuss some challenges and research directions.

**\*corresponding author:** Regina Braga – regina.braga@ufjf.br

1  **Introduction**

Agribusiness is one of the economic sectors that is strategic to worldwide. Its growth is critical and can bring benefits to the world population. One of the strategies to promote the sector is the use of data analysis techniques. According to [21], agribusiness still lacks solutions that allow the consumer market analysis to improve the sector (ex: product adequacy, sales, and cost reduction).

Traditional market research is time-consuming, expensive, and, at times, incomplete and without representation, considering the different characteristics of the consumer profile and the challenges of the agribusiness sector. Dealing with multiple stakeholders in market research projects often makes quality data collection and communication reliable, low-cost, complex, and challenging. Thus, studies carried out at Embrapa Gado de Leite [23, 22, 21, 30] point to opportunities for new strategies that could provide more significant margins in the market.

The ability of organizations to react to changes imposed by the market is directly related to the absorption of information and generation of knowledge applied to the organizations' processes. In this vein, smart tools are increasingly needed in modern industry to generate new knowledge [1]. This also applies to Agribusiness domain. Understanding consumer trends, perceptions, or preferences, economically and efficiently, is the focus of consumer science and manufacturing industries [13].

When treating large volumes of data, intelligent tools have been using big data analytics, allowing the discovery of correlations and knowledge derivation [2]. The emergence of Online Social Networks (OSN), such as Twitter, Facebook, and Instagram, provided the researchers ´community the access to voluntary information that was previously unattainable and demanded a high access cost.

In this context, the use of crowdsourcing stands out. Crowdsourcing has become an important component in several research domains such as climate change [25], natural disaster preparedness and monitoring [26, 27], conservation science [28] and urban sustainability [29]. The term refers to data collected and made available to researchers by non-professional people or organizations [32]. In opportunistic crowdsourcing, non-professionals generate data. It is collected and shared by uploading information to web-based social networking sites. In such applications, users are producers and send information as web content to be used by researchers for purposes other than those intended by users.

Social Media Analysis (SMA) and social networking sites (e.g., Twitter, blogs, or forums) can be critical to achieving a better understanding of human interactions and to defining new business models for organizations and environmental management [40]. Gohfar Khan [33] defines SMA as "the art and science of extracting valuable hidden insights from large amounts of semi-structured and unstructured social media data to enable informed and insightful decision-making". SMA is concerned with developing and evaluating tools to collect, monitor, analyze, and visualize social media data to extract relevant information and patterns [31]. Thus, SMA is a growing area that encompasses a variety of modeling and analytical techniques (e.g., sentiment analysis, topic modeling, image analysis, and others) [33].

From a structural point of view, OSNs are complex structures composed of vertices, usually representing users, and edges, which represent some form of relationship between the vertices. The study of the different types of relationships between the vertices of OSNs is called Social Network Analysis (SNA) [30]. SNA is generally used to analyze a social network graph to understand its theoretical connections and properties and identify the relative influence of the network vertices. Thus, it allows modeling the dynamics and growth of the OSN, i.e., predicting new connections, detecting communities and their associated influences, network density, among others [34]. This type of analysis is commonly used to help carry out targeted marketing campaigns, such as those most likely to buy a product.

From a business perspective, OSN represents an opportunity to reach a significant audience for market surveillance [4]. In this sense, the use of SNA and SMA contributes to information that can only be perceived from specific analyses. For example, detecting the most influential user on the network and evaluating their classified comments can mean finding the user most likely to reach a targeted marketing audience when presenting the product or reversing the lack of information to help a company advertise its products [6]. In this sense, Agribusiness can benefit from these analyzes. Thus, this work proposes a Systematic Literature Mapping (SLM) to find more accurate and representative directions for Agribusiness marketing, which require less time and allows a forecast of market trends, such as SNA/SMA techniques. In other words, this work will explore how the use of SNA and SMA can support Agribusiness.

According to [5], the Systematic Literature Mapping (SLM) is an overview of primary studies on a specific topic that aims to identify subtopics that need more primary studies. This type of study uses a protocol, following methodological steps to make the results more reliable. Thus, this mapping focuses on research associated to the use of crowdsourcing in activities related to agribusiness, which use as the main element of analysis data from online social media, in other words, "websites and applications that allow users to create and share content or participate in social networks" [38].

The research problem addressed in this paper is the need to support agribusiness in consumer market analysis to improve the sector. To tackle this problem, we investigate new strategies to improve marketing in Agribusiness. Therefore, the main contribution of this paper is a SLM that point research directions for Agribusiness marketing, which require less time and allows a forecast of market trends. We defined the following Research Question (RQ): "*How can the use of SNA and/or SMA support marketing improvement in Agribusiness?*".

For presenting the results of our investigation, this article is organized as follows, including this introduction. Section 2 presents related works to explore other literature reviews on the same subject. Section 3 highlights the research method. In Section 4, the results and discussion are presented in detail. Section 5 shows the final considerations.

## 2 Related Work

This section highlights works that present systematic reviews and mappings, relating to agribusiness and social networks. Before the SLM execution, a search was performed in March 2022, looking for other literature reviews dealing with the same topics. Therefore, we retrieved the set of secondary studies composed of 20 works that were analyzed and went through the reading the title and abstract of each one. The selected secondary studies (30%) represent works whose context is social media, whether applied to agribusiness or just cited as an example of an application domain.

In the work presented in [35], the authors present an overview of sentiment analysis in various domains. The authors addressed and identified gaps, presenting guidelines for future research in the area. One of the areas that presented opportunities mentioned by the authors is agriculture. An evaluation was also carried out comparing metrics of traditional methodologies to the approaches using a Twitter dataset, such as lexical approach, ontologies, and machine learning. As a result, the authors concluded that ontologies, vector machine support and term frequency achieved high precision and provided better results. In [36] the authors discuss the current challenges of machine learning use in domains to help decision-makers plan their actions. With 79 papers analyzed, most of them use machine learning in the industrial sector (65%), with the agricultural sector representing 5% of the total. Thus, we can consider that the agricultural sector needs more research to improve the efficiency of data analysis.

In [37], the authors present a systematic literature review that emphasizes social networks' importance in climate change interventions. The authors concluded that more studies focusing on social networks are needed. These studies can accelerate the adoption of climate-smart farming practices, helping farmers implement adaptive practices related to climate change and decision making. In another article [39], the authors discuss the main aspects of social network analysis applied to the investigation of the social life of animals in the zoo, considering that social relationships between animals influence how animals interact with the resources provided and how space is used. The paper presents a review of how SNA can be used to assess the social behavior of animals and highlights directions for future research. The authors conclude that the use of SNA can directly impact management decisions and help maintain an animal welfare standard.

In [46] the authors explore the challenges of using decision support systems in Agriculture 4.0. This paper uses the systematic literature review technique to retrieve representative decision support systems, including their applications in climate change adaptation, water resources management, and food management. Based on the result of an evaluation that was carried out on these decision support systems, challenges are detected, suggesting development trends, and demonstrating potential improvements for future research. The paper does not discuss data collection methods in online social media and does not list the techniques used in these decision support systems.

To the best of our search efforts, based on the secondary studies highlighted above, we did not find a mapping or systematic review that analyzes the solutions that use social media, such as OSN, to support agribusiness. These initial literature reviews motivated us to taggle this research gap and merge these three topics of interest: Agribusiness, SNA and SMA. Thus, our mapping study was conducted to provide state of the art in using SNA and SMA to support marketing in Agribusiness. Therefore, this SLM's main contribution is identifying, classifying, and analyzing agribusiness research using OSN. We also aim to reveal the state-of-the-art research dealing with OSN and/or SMA to support agribusiness.

## 3 SLM Methodology

According to [5], an evidence-based solution could highlight the need to use secondary studies to investigate and gather evidence on a specific topic. Through a systematic mapping, we identified the importance of analyzing the use of OSN with SMA and SNA, to assist in marketing in agribusiness.

## 3.1 Searching Strategy

We adopt the hybrid SLM method presented by [8] in this work. In this method, the authors propose four strategies that combine database searches in digital libraries with backward snowballing (BS) and forward snowballing (FS) iterative (BS*FS), parallel (BS||FS), or sequential BS+ FS and FS+BS. In addition, the authors carry out a comparative evaluation of traditional digital libraries to find the database that has the most significant performance results. The authors considered the Scopus database the most consistent digital library in terms of accuracy. In addition, the library integrates other digital libraries into its search method, increasing the search reach. However, it is necessary to complement the library with the snowballing process.

The hybrid strategy adopted in this mapping is Scopus + BS||FS. In this strategy, an initial set of papers is obtained through Scopus. Then BS and FS are performed in parallel on the same initial set. In other words, articles obtained by BS are not subject to FS and vice versa. We introduced this strategy to increase accuracy without compromising recall [14].

## 3.2 Research Questions

The following research question (RQ) was formulated to conduct the systematic mapping: "*How can the use of SNA and/or SMA support marketing in Agribusiness?*". This RQ aimed to investigate the techniques used and where these solutions are applied. Thus, we can understand how the solutions can help in decision-making. This RQ was associated with four secondary questions, as shown in Table 1.

**Table 1.** Secondary Research Questions.

| Identifier | Research Question | Goal |
|---|---|---|
| **RQ1** | What are social networks used for data collection in the SNA/SMA agribusiness research community? | Find which social networks are relevant data sources for the SNA/SMA survey in agribusiness. |
| **RQ 2** | What are the SNA/SMA analysis techniques used in agribusiness studies? | Identify SNA/SMA techniques commonly used for analyzes performed in agribusiness (e.g., sentiment analysis, influence analysis, textual analysis, among others). |
| **RQ 3** | What are the evaluation metrics? | Highlight which evaluation metrics were used to verify the proposed solution (e.g., accuracy, F1-score, number of publications, among others). |
| **RQ 4** | Which subdomains of the agribusiness are studied? | Find where researchers apply their approaches in the context of agribusiness (e.g., milk and dairy products, food security, urban agriculture, among others). |

**Table 2.** PICOC

| PICOC | Description |
|---|---|
| Population (P) | SNA and SMA solutions |
| Intervention (I): | Agriculture, Agribusiness |
| Comparison (C): | Not applicable |
| Outcome (O): | Approach |
| Context (C): | Crowdsourcing, OSN |

From the research questions, we used the PICOC method [7] to define the scope of work and the terms used in the search string, as illustrated in Table 2. The search string, the main terms of the RQs, synonyms, and acronyms were specified and validated with the assistance of an agribusiness expert and control articles [13, 15, 18, 41]. The identified terms form the following search string: ((("online social network" OR "OSN" OR "social network" OR "social media") AND ("analysis" OR "approach" OR "architecture" OR "analytics")) OR "SNA" OR "SMA") AND ("agriculture" OR "agri-food" OR "agronomy" OR "agribusiness" OR "agro-industry" OR "agricultural business" OR "dairy products"). Two external researchers revised the protocol. Furthermore, we only cover the SNA/SMA studies applied to agribusiness. From the string, we obtained 223 publications in the period 2017 – 03/2022. The search filters used are detailed in Section 3.4.

### 3.3 Inclusion and Exclusion Criteria

Due to the large number of documents retrieved by the digital library (Scopus), we used the inclusion and exclusion criteria to select only potentially relevant articles. These conditions have been applied to all documents returned from Scopus. Then, we applied the criteria to eliminate papers not related to the goals of this mapping. The inclusion and exclusion criteria used in this work are shown in Table 3. We used the Parsifal[1] tool to support the mapping execution.

The selection of studies was performed at two levels: (i) a new search was performed, adding the inclusion criteria as filters in the advanced search by Scopus; and (ii) a reviewer guided by the exclusion criteria performed a selection considering the title and abstract. It was verified whether both make explicit reference to social media or social networking services in agribusiness. Records at this level were kept when there was doubt about their relevance – just reading the title and abstract were not sufficient for the evaluation. As a result, additional sections of the articles were read.

Table 3. Inclusion and Exclusion Criteria

| Inclusion Criteria | Description |
|---|---|
| 1 | Search by title and abstract. |
| 2 | Studies published between 2017 - 03/2022. |
| 3 | Open access studies. |
| 4 | The article was published in a Journal or Conference. |
| 5 | Papers are written in English. |

| Exclusion Criteria | Description |
|---|---|
| 1 | SNA and SMA studies that do not have OSN as a data source. |
| 2 | Studies that are not applied to the agribusiness sector. |
| 3 | Books, sections, and book chapters. |
| 4 | Reviews papers. |

---

[1] https://parsif.al/

## 3.4 Quality Assessment

According to the guidelines proposed in [5], researchers can develop a quality assessment for primary studies. The evaluation serves as a guide for interpreting the results. Each of the articles selected after the inclusion and exclusion criteria was evaluated considering the quality Assessment Questions (AQ). These questions, presented in Table 5, were developed based on [9] and [10]. For each question, we assigned the value 1 if the answer was "yes", no value was assigned if "no" was answered, and 0.5 value if the answer was "partially".

We established a predefined questionnaire (Table 4 presents details) and the AQ questions (Table 5) and highlighted information such as the social media source used and the evaluation metrics to answer the questions. The publications were categorized and tabulated according to the questions, extracting their information related to the questionnaire. The questionnaire can be accessed in [2]. This technique helped detect and validate the data extraction results and settle any discrepancies.

**Table 4.** Questionnaire details.

| Questions | Example answer |
| --- | --- |
| What is the name of the paper? | #Eggs: social and online media-derived perceptions of egg-laying hen housing |
| Does the study use SNA and/or SMA? | SMA |

---

[2] https://forms.gle/Xb56pN8ybcSPYXbu7

| | |
|---|---|
| Which SNA or SMA technique is used? | Sentiment Analysis; Age and Gender Estimation; Time Analysis |
| What is the most used evaluation metric? | Statistical methods (sum, average, percentage, frequency) |
| What is being analyzed? | Eggs |
| Which social media site is used? | Netbase Analytics Platform |

Table 5. Quality Assessment Questions.

| Assessment Questions | Description |
|---|---|
| AQ1 | Is the purpose of the study clear? |
| AQ2 | Are the uses of the techniques justified and clearly described? |
| AQ3 | Are data collection methods adequately described? |
| AQ4 | Is the collected data adequately described? |
| AQ5 | Do the authors discuss limitations, threats to validity, and reliability of the study results? |
| AQ6 | Are research questions answered clearly? |
| AQ7 | How clear is the interpretation of the data collected and the conclusions in the text? |

## 3.5 Results

Based on the inclusion criteria (first level), 223 publications were returned from 2017 until March 2022. Figure 1 shows an increasing number of articles published between 2017 and 2020. This growth can be due to the growing popularity of OSNs over the years, which generated the researcher´s interest. In 2020 there were 3.96 billion active users on OSNs worldwide, and in 2017 there were only 2.79 billion active users – an overall increase of 41.63%[3]. However, in relation to the number of articles published, the year 2021/2022 was lower than 2020. A fact that may be related to this decrease is the pandemic generated by the COVID virus that started in 2020. According to [47] the searches for studies related to the disease have increased.

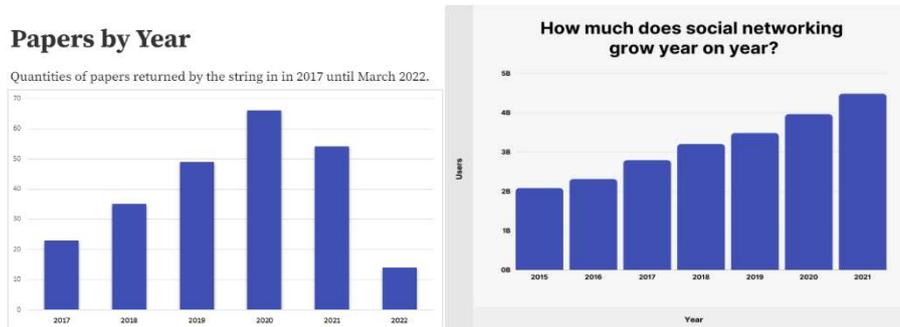

**Fig. 1.** (a) Distribution of publications per year and (b) Global social media growth rates per year by backlincko website.

To determine study eligibility, all publications that used the identified social media sources, based on the definition given in [38], were considered, including blogs, news, and other user content sharing sites (e.g., online forums). At the second level, described in Section 3.3, few studies considered OSN a collection source for SNA and SMA in agribusiness. Considering publications that exclusively use SNA or SMA (~23%), ~69% used traditional data collection techniques, such as questionnaires and interviews, using OSN as a means of communication. Others (~2%) did not use social media to collect data. They collect data through direct questionnaires. Therefore, these studies were discarded according to exclusion criterion 1 shown in Table 2. Finally, there were 12 papers that were selected as a set of studies directed to the BS||FS.

---

[3] Accessed in 07/21/2021: <https://backlinko.com/social-media-users>

A support tool[4] configured to carry out the process using Scopus as a database was used in the snowballing process. The Forward Snowballing (FS) process execution resulted in 13 papers, and the Backward Snowballing (BS) process in 408 papers. After applying the inclusion criteria, 12 papers remained in FS set and 30 in BS set. However, no papers were left in FS and BS when using the exclusion criteria. Thus, out of 223 articles, only ~5% (12 papers) remained in the final set. This significant reduction can be explained by the number of false-positive studies captured in Scopus through the word "agriculture" and its variations in the search string. The process can be seen in Figure 2.

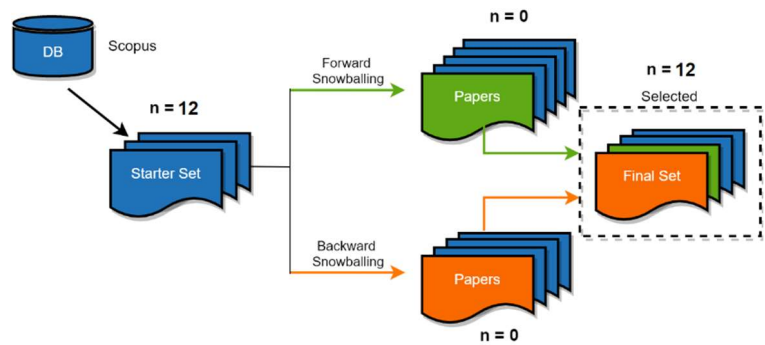

**Fig. 2.** Hybrid methodology structure (Scopus + BS||FS)

---

[4] https://github.com/JoaoFelipe/snowballing

### 3.6 Data Extraction and Analysis

The papers selected were analyzed, and the data extracted were used to answer the research questions.

**RQ1. What are social networks used for data collection in the SNA/SMA agribusiness research community?** From the questionnaire used in the Quality Assessment, we separated the social media sources for data collection to answer this question. Through this list, the number of studies (~67%) that use data from a single social media source and the number that use more than one source (~33%) were identified. We can analyze that most researchers prefer to use only one social media source for their research. According to the extracted data, Twitter (~67%) was the most used source, while Youtube, Facebook, WeChat and Sina Weibo were the least used (~8%). The last two OSNs are popular in China and are growing[5]. According to Statista[6] ranking of the most popular OSNs worldwide in January 2022, WeChat and Sina Weibo are among the top ten. Being an OSN used for opinions and considering that short texts are easily mined and processed by its free API, Twitter is the most popular among the research community. Furthermore, it was also identified that some studies (~27%) consider the use of social media analysis platforms such as Netbase[7] (~18%), which has several ONS as a data source (e.g., Twitter, Reddit, blogs, forums, and more), LikeAlyzer[8] (~8%) for Facebook, Twitonomy (~8%) for Twitter and Meltwater[9](~8%), which also has several OSN as a social media source. These analyzes can be seen in Figure 3, which illustrates the results obtained for the analysis of RQ1, showing a ranking of the most used OSNs among the community of researchers in agribusiness.

---

[5] Accessed at 02/04/2022: <https://www.statista.com/statistics/255778/number-of-active-wechat-messenger-accounts/>

[6] Accessed at 02/04/2022: <https://www.statista.com/statistics/272014/global-social-networks ranked-by-number-of-users>

[7] https://netbasequid.com/

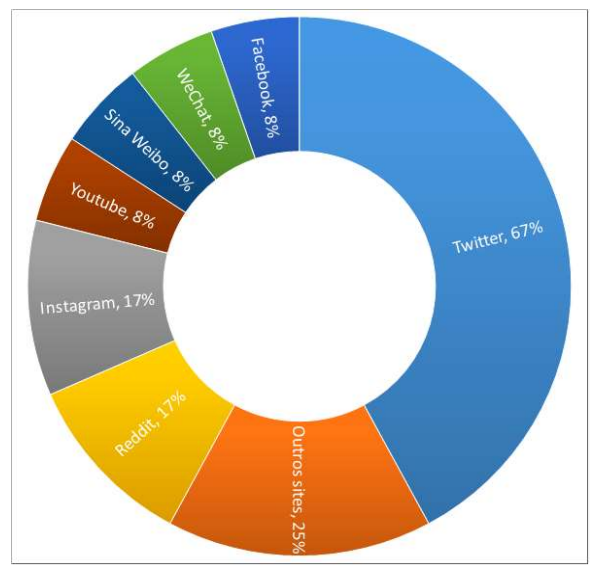

**Fig. 3.** Popularity of social media sources among researchers.

**RQ2. What are the SNA/SMA analysis techniques used in agribusiness studies?**

We investigated the studies considering which techniques were used to analyze media and social networks to answer this question. We identified that studies using SNA (~25%) were the minority. SNA was used in studies as the sole means of analysis (~8%) or combined with SMA (~17%). In other words, studies that consider SNA tend to consider the use of SMA to solve problems. These studies use textual analysis of the collected data and perform **a topological analysis** of the most cited words. Topological analysis was the only identified SNA method through the form used to model a keyword network and find trends [11, 12]. It is possible to observe that the two works that used this type of analysis also used textual analysis. Both works first use textual analysis to find the most frequent keywords. Then, they use topological analysis to understand the relationships that one keyword has with another. As for the SMA methods, the following were identified: (i) **time analysis**, where a timeline of social media is made, identifying the number of publications during a period [13, 15, 43, 45]; (ii) **textual analysis**, where social media keywords were studied using natural language processing (NLP) [11, 12, 19, 45]; (iii) **statistical analysis**, where the studies used hypothesis tests, means and percentages [16, 17, 18, 42]; (iv) **sentiment analysis**, which aims to identify and extract subjective information from social media by combining NLP and machine learning techniques to assign weighted emotional scores [13, 15, 20, 42, 43, 45]; (v) **geographic analysis**, where media upload coordinates are used to map their geographic locations [43, 45]; and finally (vi) **demographic analysis**, where age and gender of the social media user are estimated using their first and last name as input [13]. Figure 4 illustrates this analysis and shows a radar graph of the popularity of the SNA and SMA techniques. In the graph, it is possible to identify in the geographic, topological, and demographic analysis that these areas deserve attention. These areas can be better explored, as few works have been identified.

---

[8] https://likealyzer.com/

[9] https://www.meltwater.com/

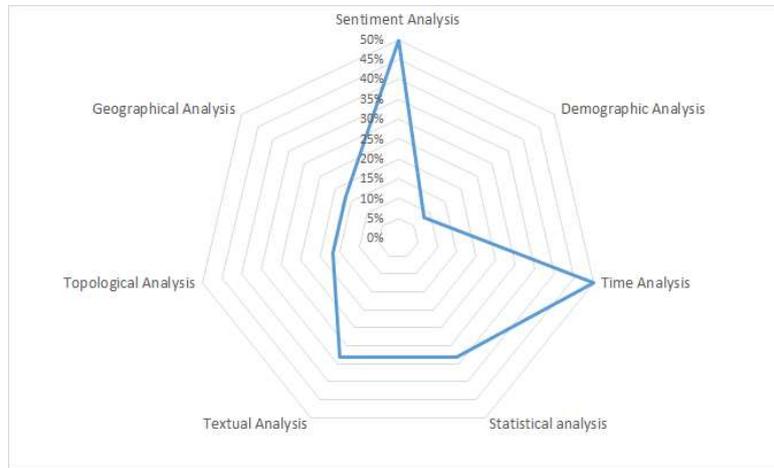

**Fig. 4.** SNA and SMA techniques popularity radar.

**RQ3. What are the used evaluation metrics?** We analyzed the selected papers to extract the researcher's metrics to understand how evaluations of the methods proposed in primary studies were carried out. Most articles (~83%) [12, 13, 15, 16, 17, 18, 19, 42, 43, 45] used statistical methods such as total publications, means, and percentages to describe the results and evaluate the proposal. However, none of them had an analysis dedicated to evaluating the proposed method. Only one of the primary studies had a section describing the proposal evaluation [20]. This study used the following metrics: precision, recall, F1-score, and Area Under Curve (AUC). These metrics were used to validate the machine learning models to classify media into a sentiment. Finally, only one study of the final set did not use any method to evaluate the proposed approach [11].

**RQ4. Which segments of the agribusiness sector are studied?** We analyzed the studies to understand the agribusiness subsectors where the SNA and SMA techniques are applied. As a result, we verified that each work considered a specific segment in agribusiness. In [13], for example, the authors studied the consumers' view on the production system for **eggs and laying hens**. The authors also exemplify how monitoring OSN can help decision-makers manage agribusiness marketing food systems through the proposed method. In [15], the authors analyzed **agricultural markets** over 27 months, searching for agricultural market-related keywords. They complete valuable insights about agricultural markets, including the public's view of the livestock industries and the risk of zoonotic diseases. In [18], the authors examine the activity engagement of companies in the **olive oil sector** in the OSNs and verify if the organic and non-organic operators have differences. According to the authors, organic food products in Spain face commercial problems due to some factors, such as the considerable price differential between organic products and their conventional equivalents.

The study reveals statistically significant differences in the engagement and use of OSNs by non-organic and organic operators, being the second most active in the OSNs used in the study. In [16], the authors combine data from 13 sites in 11 low-income countries to study how various social capital scales relate to **household food security** outcomes among smallholders. The authors conclude that social network theory correlates household food security with multiple social capital scales, both within and outside the household. This social capital can be either a link (within groups) or a bridge (between groups) with different implications for how the structure of social capital affects food security. The article [12] presents the first content analysis on the Czech Twitter OSN in the context of **agriculture in general**. The authors found significant focus in tweets about biofuels, the rapeseed plant, and politics.

Furthermore, they conclude that robot accounts created a significant proportion of tweets. In [17], the authors share their experience using the WhatsApp platform for communication and data collection to monitor and evaluate the sweet potato value chain. The article [11] drew only on data from public newspaper reports and a sample of the social networks used by **urban food** networks in Bristol - a city with a well-developed urban agriculture movement - to explore how activists in urban agriculture food were related to the media during 2015. The authors intended to inform debates on urban agriculture and contribute to discussions on its growth. In [19] the authors suggest understanding the influence of COVID-19 on China's **agricultural economy**. Thus, it was possible to gain necessary insights while the virus continues to spread worldwide. In [42], the authors address the problem caused by wild pigs for agriculture and the environment. Through SMA, the authors find evidence of a lack of information on best practices for safety, such as the risk of zoonotic diseases caused by wild pigs. In addition, they describe the importance of understanding the influence of social media on people and opportunities for management agencies, such as messages in public health campaigns. In [43] the authors explore Artificial Intelligence (AI) in agriculture. Based on SMAs, the authors conclude that AI techniques in agriculture were positive. In the work presented in [45], the authors focus on how the use of SMA can support government authorities to predict damages related to the impacts of natural disasters in urban centers. The study uses SMA in Twitter crowdsourced data using the keywords "Disaster" and "Damages". The study's methodological approach employs the social media analysis method and performs sentiment analysis and textual content of Twitter messages.

Finally, the paper [20] uses public opinion in OSN to determine whether **Smart Agriculture** or **Agriculture 4.0** is implemented in Indonesia. Thus, answering **RQ4**, ten segments were identified: eggs and laying hens, agricultural markets, olive oil, family food security, agriculture in general, sweet potato, urban agriculture, agricultural economy, intelligent agriculture, wild pigs, impacts of natural disasters in urban centers and AI in agriculture. In addition, several agricultural segments have not been explored in the OSN context and analysis techniques, such as milk and its derivatives, such as cheese, yogurt, and other commodities.

## 4   Research Directions

As social media and network analytics evolve in agribusiness, it is possible to identify various social media sites and analytics platforms used. The most popular, if not the most representative, social media site is OSN Twitter.

We observed that social media are primarily collected using Twitter as a source, which is used in ~64% of the selected studies. Thus, the Twitter platform is the most used by researchers. In general, data from OSNs are the most used, either as a single source of collection, in conjunction with another OSN, or through analysis platforms.

The most used analysis technique in agriculture is the machine learning technique in sentiment analysis, which is used in ~50% of the selected studies. However, we find increasing activity in both time-based techniques and statistics.

Our evaluation metrics observed that precision, recall, and accuracy score are the evaluation metrics most used by the selected studies.

After analyzing the selected studies, we identified some research gaps that are important to be investigated. From the SLM results, it was possible to conclude that the data collected from the Twitter platform is the most used dataset in SNA and SMA applied to agribusiness compared to other social media platforms. This preference for Twitter is mainly due to the ease of access. However, with the growth in importance of other social media and mechanisms for extracting information concerning different media types such as photos and videos, it is necessary to advance research. This deficiency limits the generalization of the results obtained through the analyzes.

Thus, we conclude that research directions should focus on the use of multiple OSNs combined with the application of information extraction and analysis techniques in the various segments of agriculture, especially the underexplored segments, for example milk and dairy segments. We can also explore techniques such as topological analysis and use of semantics, which can assist in extracting information from new media like augmented reality.

Integrating information from various OSNs and the analysis of this data, correlating findings, and directing marketing strategies is an important research direction. The knowledge discovery from this information correlation can help discover implicit relationships between data. In addition, the use of traceability techniques to verify the integrity of information is also important.

All these research directions are important and should be investigated in future research. In this sense, this mapping promotes strategies based on new techniques for acquiring knowledge to leverage agribusiness. In addition, the SLM results also assist researchers in a better understanding of research directions related to the use of media and social network analysis techniques in OSNs applied to agribusiness.

Considering our RQ: "*How can the use of SNA and/or SMA support marketing improvement in Agribusiness?*", the results of SLM can provide directions to follow. The investigation using multiple OSNs is a path to follow. Machine Learning techniques are also pointed as a technique to be used.

Therefore, as a result of our investigation, we consider that the following techniques should be investigated: i) use of multiple OSNs, combining their results in some way, ii) use of intelligent data analysis techniques, emphasizing ML techniques, but also investigating other techniques such as semantic and structural analysis, given the structure of an OSN, iii) provide adequate visualization mechanisms to leverage the commercialization of specific products.

## 5 Threats to Validity

This systematic mapping aimed to provide an overview of the literature regarding the use of OSN, identifying, categorizing, and analyzing SNA and SMA solutions in the agricultural domain. However, some threats to validity and limitations can influence the results.

As a threat to the **construction validity**, we can mention the search string (see Subsection 3.3). We defined the main terms of the RQ, synonyms, and acronyms considered adequate to make the string as comprehensive as possible. However, some terms may not have been considered in the string. For that, several tests were carried out with the terms, and as we carried out the searches, versions were generated to decide on the best string. In addition, experts reviewed, and control papers were used. As a result, this threat was mitigated.

All formulated conclusions, and results found in this SLM, have traceability. However, biased data extraction from selected articles may threaten the **conclusion validity**. In other words, we may have been including papers into the final selection set, but they are not suitable, thus being a false positive. We used a predefined form of data extraction to mitigate this threat (see Subsection 3.5).

Regarding threats to internal validity, the SLM selection process (see Subsection 3.1) was conducted by only one researcher, and it may be difficult to include all relevant publications in the research. Furthermore, it isn't easy to ensure that all topic-related concepts and relevant articles have been included in this study, despite care and effort. However, the snowballing technique was used, which helped include new relevant works.

## 6 Final Remarks

This paper presents the research directions in data analysis practices and tools from social media sources, such as OSN, related to agribusiness. We started with 26,551 publications returned by the search string in Scopus that was filtered and went through an evaluation process during the SLM. As a result, twelve primary studies were selected.

Applications involving summarizing opinions, spotting trends, and sentiment analyses are the most common techniques these tools offer. This study focuses on the analytical approaches and techniques employed to develop such applications and the social media sources used to collect the data. We identified that a combination of different analysis methods such as temporal, sentiment, and topological analysis are often used together to achieve a specific analysis objective more accurately and with an overview of agribusiness.

As a contribution, this mapping study brings state of the art in using SNA and SMA to leverage agribusiness. The study identified online social media sources, platforms, and techniques for analyzing information, assessment metrics, and the agricultural segments covered by the primary studies.

Through the SLM, it was noticed that there was a lack of primary studies on SNA and SMA in agribusiness. For example, in the SNA context, few works use community analysis and influence analysis techniques. In addition, one of the untreated agricultural segments, which has significant economic importance, is milk and dairy products. Thus, there is a need to assist marketing in the agricultural sector using analytical methods and opportunistic crowdsourcing.

The results obtained by the primary studies selected in this SLM, together with insights into online social media data analysis, reinforce how social media monitoring can complement traditional methods to inform agricultural producers and consumers about systems management and regulation of agribusiness.

As future work, it is up to the execution of this SLM to mitigate threats to validity. For example, the use of two or more researchers in the data extraction process and increased bibliographic sources for data collection. We can also reduce the threats to the construct validity of the SLM by updating the search string terms. Over the years, new terms associated with technologies not yet explored in agribusiness may emerge.

## Acknowledgments

We would like to thank the researchers who participated in the evaluation of this proposal, as well as the Brazilian Agricultural Research Corporation (Embrapa/Brazil).

## Funding

This work was partially funded by UFJF/Brazil, CAPES/Brazil, CNPq/Brazil (grant: 311595/2019-7), and FAPEMIG/Brazil (grant: APQ-02685-17), (grant: APQ-02194-18).